\documentclass[aip, jmp, showpacs, showkeys, citeautoscript]{revtex4-1}
\usepackage{amsfonts}
\usepackage{amssymb}
\usepackage{amstext}
\usepackage{amsmath}
\usepackage{latexsym}
\usepackage{color}

\newtheorem{thm}{Theorem}

  \def\ulamek#1#2{\mbox{\normalfont$\frac{#1}{#2}$}}

\begin{document}

\title[On the properties of Laplace transform originating from one-sided L\'{e}vy stable laws]
{On the properties of Laplace transform originating from one-sided L\'{e}vy stable laws}

\author{K.~A.~Penson}
\email{penson@lptl.jussieu.fr}

\affiliation{Sorbonne Universit\'{e}s, Universit\'e Pierre et Marie Curie (Paris 06), CNRS UMR 7600\\
Laboratoire de Physique Th\'eorique de la Mati\`{e}re Condens\'{e}e (LPTMC),\\
Tour 13 - 5i\`{e}me \'et., B.C. 121, 4 pl. Jussieu, F 75252 Paris Cedex 05, France\vspace{2mm}}

\author{K.~G\'{o}rska}
\email{katarzyna.gorska@ifj.edu.pl}

\affiliation{H. Niewodnicza\'{n}ski Institute of Nuclear Physics, Polish Academy of Sciences, Division of Theoretical Physics, ul. Eliasza-Radzikowskiego 152, PL 31-342 Krak\'{o}w, Poland}



\begin{abstract}
We consider the conventional Laplace transform of $f(x)$, denoted by $\mathcal{L}[f(x); p]~\equiv~F(p)=\int_{0}^{\infty} e^{-p x} f(x) dx$ with ${\rm \mathfrak{Re}}(p) > 0$. For $0 < \alpha < 1$ we furnish the closed form expressions for the inverse Laplace transforms $\mathcal{L}^{-1}[F(p^{\alpha}); x]$ and $\mathcal{L}^{-1}[p^{\alpha-1}F(p^{\alpha}); x]$. In both cases they involve definite integration with kernels which are appropriately rescaled one-sided L\'{e}vy stable probability distribution functions $g_{\alpha}(x)$, $0 < \alpha < 1$, $x > 0$. Since $g_{\alpha}(x)$ are exactly and explicitly known for rational $\alpha$, \textit{i.e.} for $\alpha = l/k$ with $l, k=1, 2, \ldots$, $l < k$, our results extend the known and tabulated case of $\alpha = 1/2$ to any rational $0 < \alpha < 1$. We examine the integral kernels of this procedure as well as the resulting two kinds of L\'{e}vy integral transformations.
\end{abstract}

\maketitle

%----------------------------------------
\section{Introduction}

The (direct) Laplace transform of the function $f(x)$, $x > 0$, is defined as
\begin{equation}\label{21Feb14-1}
\mathcal{L}[f(x); p] \equiv F(p) = \int_{0}^{\infty} \exp(-p x) f(x) dx, \quad {\rm \mathfrak{Re}}(p) > 0,
\end{equation} 
and its inverse $\mathcal{L}^{-1}$ can be formally written as 
\begin{equation}\label{21Feb14-2}
\mathcal{L}^{-1}[F(p); x] = f(x).
\end{equation}
(The condition ${\rm \mathfrak{Re}}(p) > 0$ is assumed to hold throughout the paper.) The usefulness of the Laplace transform in all the fields of science and technology does not need to be emphasized. Numerous physical applications of Laplace transform are worked out in [\onlinecite{LDebnath07}]. Many other physical and engineering applications are presented, from the computational viewpoint, in [\onlinecite{UGraf04}]. The practical use of the Laplace transform is enhanced by the existence of voluminous tables: see [\onlinecite{APPrudnikov-v4}] for the direct Laplace transform and [\onlinecite{APPrudnikov-v5}] for the inverse Laplace transform. 

We shall be concerned in this note with the extensions of two integral formulas for the inverse Laplace transform: 
\begin{equation}\label{21Feb14-3}
\mathcal{L}^{-1}[F(p^{1/2}); x] = \frac{1}{2 \sqrt{\pi x^{3}}} \int_{0}^{\infty} t \exp\left(-\frac{t^{2}}{4x}\right) f(t) dt,
\end{equation}
which is the formula 1.1.1.26, p. 4 of [\onlinecite{APPrudnikov-v5}], and 
\begin{equation}\label{21Feb14-4}
\mathcal{L}^{-1}[p^{-1/2} F(p^{1/2}); x] = \frac{1}{\sqrt{\pi x}} \int_{0}^{\infty} \exp\left(-\frac{t^{2}}{4x}\right) f(t) dt,
\end{equation}
which is the formula 1.1.1.31, p. 5 of [\onlinecite{APPrudnikov-v5}]. Compare also the closely related formulas 1.1.5.22 and 1.1.5.23, p. 8 of [\onlinecite{APPrudnikov-v4}]. For the derivation of Eqs. \eqref{21Feb14-3} and \eqref{21Feb14-4}, see Eqs. \eqref{21Feb14-13c} and \eqref{21Feb14-14f} respectively, below. 

The purpose of this work is to place Eqs. \eqref{21Feb14-3} and \eqref{21Feb14-4} in a wider context in order to derive more general Laplace inversion formulas which are not tabulated so far. The new ingredient here are well-defined integrable probability distribution functions called L\'{e}vy stable functions, and denoted by $g_{\alpha}(x)$, $x > 0$. They are defined through their Laplace transform as follows [\onlinecite{HPollard46, KAPenson10}]:
\begin{equation}\label{21Feb14-5}
\mathcal{L}[g_{\alpha}(x); p] = \exp(-p^{\alpha}), \quad 0 < \alpha < 1.
\end{equation}
See [\onlinecite{KAPenson10, Zolotaryev-b, Uchaikin-b}] for more properties of $g_{\alpha}(x)$; The right hand side of Eq. \eqref{21Feb14-5} is the so-called stretched exponential, also sometimes referred to as the Kohlrausch-Watts-Williams function, see [\onlinecite{GDattoli14}] and references therein. The simplest case of $g_{\alpha}(x)$ is $g_{1/2}(x)$, a historically first known L\'{e}vy stable function [\onlinecite{JPKahane95}], which reads: 
\begin{equation}\label{21Feb14-6}
g_{1/2}(x) = (2\sqrt{\pi} x^{3/2})^{-1} \exp[-1/(4x)], \quad x>0,
\end{equation}
also known as L\'{e}vy-Smirnov function. 

The paper is structured as follows: in Sec. II we present our main results, in Sec. III we discuss further properties of integral kernels of Sec. II and in Sec IV we present conclusions and discussion.

\section{Derivation of main results}

As the first step we shall rewrite Eqs. \eqref{21Feb14-3} and \eqref{21Feb14-4} using the function $g_{1/2}(x)$ of Eq. \eqref{21Feb14-6}. It is easy to see that 
\begin{equation}\label{21Feb14-7}
\mathcal{L}^{-1}[F(p^{1/2}); x] = \int_{0}^{\infty} \frac{1}{t^{2}} g_{1/2}\left(\frac{x}{t^{2}}\right) f(t) dt, 
\end{equation}
as well as,
\begin{equation}\label{21Feb14-8}
\mathcal{L}^{-1}[p^{-1/2} F(p^{1/2}); x] = 2x \int_{0}^{\infty} \frac{1}{t^{3}} g_{1/2}\left(\frac{x}{t^{2}}\right) f(t) dt.
\end{equation}
Anticipating the sought for generalization for $\alpha \neq 1/2$ we introduce the following two positive two-variable kernels for $t, x > 0$:
\begin{align}\label{21Feb14-9}
M_{\alpha}(t, x) & = \frac{1}{t^{1/\alpha}} g_{\alpha}\left(\frac{x}{t^{1/\alpha}}\right), \\ \label{21Feb14-10}
N_{\alpha}(t, x) & = \frac{x}{\alpha t} M_{\alpha}(t, x) = \frac{x}{\alpha t^{1+1/\alpha}} g_{\alpha}\left(\frac{x}{t^{1/\alpha}}\right),
\end{align}
which in turn define the following two L\'{e}vy integral transformations:
\begin{equation}\label{21Feb14-11}
\tilde{f}_{\alpha}(x) = \int_{0}^{\infty} M_{\alpha}(t, x) f(t) dt = \int_{0}^{\infty} \frac{1}{t^{1/\alpha}} g_{\alpha}\left(\frac{x}{t^{1/\alpha}}\right) f(t) dt,
\end{equation}
and
\begin{equation}\label{21Feb14-12}
\bar{f}_{\alpha}(x) = \int_{0}^{\infty} N_{\alpha}(t, x) f(t) dt = \frac{x}{\alpha} \int_{0}^{\infty} \frac{1}{t^{1+1/\alpha}} g_{\alpha}\left(\frac{x}{t^{1/\alpha}}\right) f(t) dt.
\end{equation}
We tacitly assume that the integrals in Eqs. \eqref{21Feb14-11} and \eqref{21Feb14-12} are convergent. Observe also that $\tilde{f}_{1/2}(x) = \mathcal{L}^{-1}[F(p^{1/2}); x]$ and $\bar{f}_{\alpha}(x) = \mathcal{L}^{-1}[p^{-1/2} F(p^{1/2}); x]$. The forms of Eqs. \eqref{21Feb14-9} and \eqref{21Feb14-10} may look at hoc at this stage but their relevance will become evident later on. 

In the following, assuming $\mathcal{L}[f(x); p] = F(p)$, we calculate the Laplace transforms of $\tilde{f}_{\alpha}(x)$ and of $\bar{f}_{\alpha}(x):$
\begin{align}\label{21Feb14-13}
\mathcal{L}[\tilde{f}_{\alpha}(x); p] & = \int_{0}^{\infty} e^{- p x} \left[\int_{0}^{\infty} \frac{1}{t^{1/\alpha}} g_{\alpha}\left(\frac{x}{t^{1/\alpha}}\right) f(t) dt\right] dx \\ \label{21Feb14-13a}
& = \int_{0}^{\infty} f(t) \left[\int_{0}^{\infty} e^{- p x} \frac{1}{t^{1/\alpha}} g_{\alpha}\left(\frac{x}{t^{1/\alpha}}\right) dx\right] dt \\ \label{21Feb14-13b}
& = \int_{0}^{\infty} f(t) \left[\int_{0}^{\infty} e^{- p t^{1/\alpha} y} g_{\alpha}(y) dy\right] dt \\ \label{21Feb14-13c}
& = \int_{0}^{\infty} e^{- p^{\alpha} t} f(t) dt = F(p^{\alpha}),
\end{align}
compare [\onlinecite{KGorska12}]. Analogously:
\begin{align}\label{21Feb14-14}
\mathcal{L}[\bar{f}_{\alpha}(x); p] & = \int_{0}^{\infty} e^{- p x} \left[\frac{x}{\alpha} \int_{0}^{\infty} \frac{1}{t^{1+1/\alpha}} g_{\alpha}\left(\frac{x}{t^{1/\alpha}}\right) f(t) dt\right] dx \\  \label{21Feb14-14b}
& = \int_{0}^{\infty} \frac{1}{\alpha t^{1+1/\alpha}} f(t) \left[\int_{0}^{\infty} x e^{- p x} \frac{1}{t^{1/\alpha}} g_{\alpha}\left(\frac{x}{t^{1/\alpha}}\right) dx\right] dt \\  \label{21Feb14-14c} 
& = \int_{0}^{\infty} \frac{1}{\alpha t} f(t) \left[\int_{0}^{\infty} t^{1/\alpha} u e^{- p t^{1/\alpha} u} g_{\alpha}(u) du\right] dt  \\  \label{21Feb14-14d}
& = \int_{0}^{\infty} \frac{1}{\alpha t} f(t) \left[ - \frac{d}{d p} \int_{0}^{\infty} e^{- p t^{1/\alpha} u} g_{\alpha}(u) du\right] dt \\  \label{21Feb14-14e}
& = \int_{0}^{\infty} \frac{1}{\alpha t} f(t) \left(-\frac{d}{d p} e^{- t p^{\alpha}}\right) dt  \\[0.4\baselineskip]  \label{21Feb14-14f}
& = p^{\alpha-1}\int_{0}^{\infty} e^{- t p^{\alpha}} f(t) dt = p^{\alpha - 1} F(p^{\alpha}).
\end{align}
In Eq. \eqref{21Feb14-13a} we have applied a simple change of variable and in Eq. \eqref{21Feb14-13c} we have used Eq. \eqref{21Feb14-5}, see [\onlinecite{KGorska12}]. Similarly, the change of variable was applied in Eq. \eqref{21Feb14-14c} and Eq. \eqref{21Feb14-5} was used in Eq. \eqref{21Feb14-14e}. The above results can be summarized in the following statement: 
\begin{thm}
If $\mathcal{L}[f(x); p] = F(p)$ and $0 < \alpha <1$, then with Eq. \eqref{21Feb14-11} 
\begin{equation}\label{21Feb14-15}
\mathcal{L}[\tilde{f}_{\alpha}(x); p] = F(p^{\alpha}), 
\end{equation}
and with Eq. \eqref{21Feb14-12} 
\begin{equation}\label{21Feb14-16}
\mathcal{L}[\bar{f}_{\alpha}(x); p] = p^{\alpha-1} F(p^{\alpha}).
\end{equation}
\end{thm}
%Evidently, Eqs. \eqref{21Feb14-15} and \eqref{21Feb14-16} hold under a tacit assumption that, for a given $f(t)$, the integrals of Eqs. \eqref{21Feb14-11} and \eqref{21Feb14-12} are convergent. 

In our opinion Eq. \eqref{21Feb14-15} is a far-reaching generalization of the formula 1.1.1.26, p. 4 of [\onlinecite{APPrudnikov-v5}], see our Eq. \eqref{21Feb14-3}, and Eq. \eqref{21Feb14-16} similarly generalizes the formula 1.1.1.31, p. 5 of [\onlinecite{APPrudnikov-v5}], see our Eq. \eqref{21Feb14-4}. Since $g_{\alpha}(x)$ are exactly and explicitly known for any rational $0 < \alpha < 1$, see [\onlinecite{KAPenson10}], Eqs. \eqref{21Feb14-15} and \eqref{21Feb14-16} are explicit for this case. For reader's convenience we reproduce below the Maple$^{\text{\textregistered}}$ procedure {\ttfamily LevyDist} which generates $g_{\alpha}(x)$ for $\alpha$ rational ($\alpha = l/k$), compare [\onlinecite{KAPenson10}]:\\
{\ttfamily LevyDist:=proc(k,l,x) simplify(convert(sqrt(k*l)\\[0.2\baselineskip] 
*MeijerG([[],[seq(j1/l,j1=0..l-1)]], [[seq(j2/k,j2=0..k-1)],[]], \\[0.2\baselineskip] 
l\textasciicircum l/(k\textasciicircum k*x\textasciicircum l))/(x*(2*Pi)\textasciicircum ((k-l)/2)), StandardFunctions)); end;}. 

\section{Further properties of integral kernels}

In addition to properties displayed in Eqs. \eqref{21Feb14-15} and \eqref{21Feb14-16} the integral kernels $M_{\alpha}(t, x)$ and $N_{\alpha}(t, x)$ possess further characteristics connected with integration over positive half-axis. We set out to evaluate the following integral of convolution type:
\begin{align}\label{24/05-1}
\int_{0}^{\infty} M_{\alpha}(t, x) M_{\beta}(y, t) dt & = \int_{0}^{\infty} \frac{1}{t^{\frac{1}{\alpha}}} g_{\alpha}\left(\frac{x}{t^{\frac{1}{\alpha}}}\right) \frac{1}{y^{\frac{1}{\beta}}} g_{\beta}\left(\frac{t}{y^{\frac{1}{\beta}}}\right) dt \\ \label{24/05-2}
& = \int_{0}^{\infty} \frac{1}{y^{\frac{1}{\alpha\beta}} u^{\frac{1}{\alpha}}} g_{\alpha}\left(\frac{x}{y^{\frac{1}{\alpha\beta}} u ^{\frac{1}{\alpha}}}\right) g_{\beta}(u) du \\ \label{24/05-3}
& = \frac{1}{y^{\frac{1}{\alpha\beta}}} \int_{0}^{\infty} \frac{1}{u^{\frac{1}{\alpha}}} g_{\alpha}\left(\frac{x/y^{\frac{1}{\alpha\beta}}}{u^{\frac{1}{\alpha}}}\right) g_{\beta}(u) du \\ \label{24/05-4}
& = \frac{1}{y^{\frac{1}{\alpha\beta}}} g_{\alpha\beta}\left(\frac{x}{y^{\frac{1}{\alpha\beta}}}\right) \equiv M_{\alpha\beta}(y, x).
\end{align}
In Eq. \eqref{24/05-2} we have performed a change of variable $u y^{1/\beta} = t$, and in Eq. \eqref{24/05-3} we have used Eq. (29) of [\onlinecite{KGorska12}]. Very similarly the following convolution-type integral reads: 
\begin{align}\label{24/05-5}
\int_{0}^{\infty} N_{\alpha}(t, x) N_{\beta}(y, t) dt & = \int_{0}^{\infty} \frac{x}{\alpha t^{1+\frac{1}{\alpha}}} g_{\alpha}\left(\frac{x}{t^{\frac{1}{\alpha}}}\right) \frac{t}{\beta y^{1+\frac{1}{\beta}}} g_{\beta}\left(\frac{t}{y^{\frac{1}{\beta}}}\right) dt \\ \label{24/05-6}
& = \frac{x}{\alpha\beta y} \int_{0}^{\infty} \frac{1}{t^{\frac{1}{\alpha}}} g_{\alpha}\left(\frac{x}{t^{\frac{1}{\alpha}}}\right) g_{\beta}\left(\frac{t}{y^{\frac{1}{\beta}}}\right) \frac{1}{y^{\frac{1}{\beta}}} dt \\ \label{24/05-7}
& = \frac{x}{\alpha\beta y^{1+\frac{1}{\alpha\beta}}} \int_{0}^{\infty} \frac{1}{u^{\frac{1}{\alpha}}} g_{\alpha}\left(\frac{x/y^{\frac{1}{\alpha\beta}}}{u^{\frac{1}{\alpha}}}\right) g_{\beta}(u) du \\ \label{24/05-8}
& = \frac{1}{\alpha\beta} \frac{x}{y^{1+\frac{1}{\alpha\beta}}} g_{\alpha\beta}\left(\frac{x}{y^{\frac{1}{\alpha\beta}}}\right) \equiv N_{\alpha\beta}(y, x).
\end{align}
In Eq. \eqref{24/05-6} we have used the change of variable $u y^{1/\beta} = t$, followed by employing again Eq. (29) of [\onlinecite{KGorska12}] in Eq. \eqref{24/05-7}. Thus Eqs. \eqref{24/05-4} and \eqref{24/05-8} express the transitivity of kernels $M_{\alpha}$ and $N_{\alpha}$ with respect to the index $\alpha$, via convolution-type integration of type Eq. \eqref{24/05-1} and \eqref{24/05-5}. They generalize Eq. (29) of [\onlinecite{KGorska12}] and are valid for any real $\alpha$, $\beta$ such that $0 < \alpha, \beta < 1$. We remark that neither $M_{\gamma}(y, x)$ nor $N_{\gamma}(y, x)$, $0 < \gamma < 1$, are symmetric with respect to $x$ and $y$.

This transitivity appears to break down for other type of convolution integrals which mix $M_{\alpha}$ and $N_{\alpha}$, namely for $\int_{0}^{\infty}M_{\alpha}(t, x) N_{\beta}(y, t) dt$. In fact the evaluation of this last integral for arbitrary $\alpha$ and $\beta$ does not appear to be possible. However, we shall demonstrate that for specific forms of $\alpha$ and $\beta$, namely for $\alpha$ and $\beta$ rational, this evaluation is possible in terms of special functions. As a first step we define the Laplace transform of the aforementioned integral through
\begin{equation}\label{24/05-10}
\mathcal{F}_{\alpha, \beta}(p, y) = \mathcal{L}\left[\int_{0}^{\infty} M_{\alpha}(t, x) N_{\beta}(y, t) dt; p\right],
\end{equation}
where, by convention, the integration variable for the Laplace transform is fixed to be $x$, and $y$ becomes a parameter. (We remind the reader that in Eq. \eqref{24/05-10} all the variables $t$, $x$, $y$ and $p$ are positive). For the moment no further assumptions about $\alpha$ and $\beta$ are made, except $0 < \alpha, \beta < 1$. Quite surprisingly, $\mathcal{F}_{\alpha, \beta}(p, y)$ can be calculated exactly:
\begin{align}\label{24/05-11}
\mathcal{F}_{\alpha, \beta}(p, y) & = \int_{0}^{\infty} e^{-px} \left[\int_{0}^{\infty} \frac{1}{t^{\frac{1}{\alpha}}} g_{\alpha}\left(\frac{x}{t^{\frac{1}{\alpha}}}\right) \frac{t}{\beta y^{1 + \frac{1}{\beta}}} g_{\beta}\left(\frac{t}{y^{\frac{1}{\beta}}}\right) dt\right] dx \\ \label{24/05-12}
& = \int_{0}^{\infty} \frac{t}{\beta y^{1 + \frac{1}{\beta}}} g_{\beta}\left(\frac{t}{y^{\frac{1}{\beta}}}\right) \left[\int_{0}^{\infty} e^{-px} \frac{1}{t^{\frac{1}{\alpha}}} g_{\alpha}\left(\frac{x}{t^{\frac{1}{\alpha}}}\right) dx\right] dt.
\end{align}
The inner integral in Eq. \eqref{24/05-12}, via the change of variable $u t^{1/\alpha} = x$ has the value $e^{-p^{\alpha} t}$, see Eq. \eqref{21Feb14-5}, and therefore
\begin{align}\label{24/05-13}
\mathcal{F}_{\alpha, \beta}(p, y) & = \int_{0}^{\infty} \frac{t}{\beta y^{1 + \frac{1}{\beta}}} e^{-p^{\alpha} t} g_{\beta}\left(\frac{t}{y^{\frac{1}{\beta}}}\right) dt \\ \label{24/05-14}
& = \frac{1}{\beta y^{1-\frac{1}{\beta}}} \int_{0}^{\infty} z e^{-p^{\alpha} y^{\frac{1}{\beta}} z} g_{\beta}(z) dz \\ \label{24/05-15}
& = \frac{1}{\beta y^{1-\frac{1}{\beta}}} \left[-\frac{d}{d a} \int_{0}^{\infty} e^{-az} g_{\beta}(z) dz\right]_{a = p^{\alpha} y^{\frac{1}{\beta}}} \\ \label{24/05-16}
& = - \frac{1}{\beta y^{1-\frac{1}{\beta}}} \left[\frac{d}{d a}\left(e^{-a^{\beta}}\right)\right]_{a = p^{\alpha} y^{\frac{1}{\beta}}} \\ \label{24/05-17}
& = p^{\alpha(\beta-1)} e^{- y p^{\alpha\beta}},
\end{align}
valid for \textit{any} real $0 < \alpha, \beta < 1$. In Eq. \eqref{24/05-13} we used the change of variable $z y^{1/\beta} = t$ and in Eq. \eqref{24/05-15} we employed again Eq. \eqref{21Feb14-5}. Eq. \eqref{24/05-17} immediately implies the equality
\begin{equation}\label{29/05-1}
\mathcal{F}_{\alpha, \beta}(p, y) = p^{\alpha-\beta} \mathcal{F}_{\beta, \alpha}(p, y).
\end{equation}
Consequently, Eq. \eqref{24/05-17} implies
\begin{align}\label{24/05-18}
\mathcal{J}_{\alpha, \beta}(x, y)\equiv\int_{0}^{\infty} M_{\alpha}(t, x) N_{\beta}(y, t) dt & = \mathcal{L}^{-1}\left[\mathcal{F}_{\alpha, \beta}(p, y); x\right] \\ \label{24/05-19}
& = \mathcal{L}^{-1}\left[p^{\alpha(\beta-1)} e^{-y p^{\alpha\beta}}; x\right].
\end{align}
According to the Laplace inversion formula 2.2.1.19 on p. 53 of [\onlinecite{APPrudnikov-v5}] Eq. \eqref{24/05-19} can be inverted if $\alpha \beta < 1$ is rational, in terms of Meijer $G$ functions [\onlinecite{APPrudnikov-v3}]. Therefore for $\alpha$ and $\beta$ rational the integral $\int_{0}^{\infty} M_{\alpha}(t, x) N_{\beta}(y, t) dt$ can be evaluated exactly.

We shall now exemplify the Laplace inversion of Eq. \eqref{24/05-19} with three pairs of rational parameters $\alpha$ and $\beta$. First we reproduce the inversion formula 2.2.1.19 of [\onlinecite{APPrudnikov-v5}] using our parametrization. It reads then, for $\alpha\beta\equiv l/k$, with $l$ and $k$ relatively prime positive integers, as
\begin{align}\label{29/05-2}
\mathcal{L}^{-1}\left[\frac{1}{p^{\alpha-l/k}} e^{-y p^{l/k}}; x\right] &= \frac{\sqrt{k}}{(2\pi)^{\frac{k-l}{2}}} l^{\frac{l}{2} - \alpha + \frac{l}{k}} x^{\alpha - \frac{l}{k} -1} G^{k, 0}_{l, k}\left(\left(\frac{y}{k}\right)^{k} \left(\frac{l}{k}\right)^{l}\Big\vert {\Delta(l, \alpha-l/k) \atop \Delta(k, 0)}\right) \\ \label{29/05-3}
& = \frac{\sqrt{k}}{(2\pi)^{\frac{k-l}{2}}} l^{\frac{l}{2} - \alpha + \frac{l}{k}} x^{\alpha - \frac{l}{k} - 1} G\left([[\,\,\,], [\Delta(l, \alpha-\ulamek{l}{k})]], [[\Delta(k, 0)], [\,\,\,]], \left(\ulamek{y}{k}\right)^{k} \left(\ulamek{l}{x}\right)^{l}\right),
\end{align}
where in Eqs. \eqref{29/05-2} and \eqref{29/05-3} for the Meijer $G$ function, denoted by $G$, we employed a traditional notation, see [\onlinecite{APPrudnikov-v3}], and a simplified notation inspired by the computer algebra systems [\onlinecite{CAS}], respectively. Above, $\Delta(k, a) = \ulamek{a}{k}, \ulamek{a+1}{k}, \ldots, \ulamek{a+k-1}{k}$, $k \neq 0$, is a special sequence. For a precise definition of Meijer $G$ function as a Mellin transform see [\onlinecite{APPrudnikov-v3}] and [\onlinecite{NP}].\\

\noindent
(a) \underline{Case $\alpha=2/3$ and $\beta=1/2$}\\
Here $\alpha\beta=1/3$, $l=1$, $k=3$, $\alpha-1/3 = 1/3$.
\begin{align}\label{29/05-4}
\mathcal{J}_{\frac{2}{3}, \frac{1}{2}}(x, y) &= \mathcal{L}^{-1}\left[p^{-\frac{1}{3}}\exp\left(-yp^{\frac{1}{3}}\right); x\right] \\ \label{29/05-5}
& = \frac{\sqrt{3}}{2\pi x^{2/3}} G^{3, 0}_{1, 3}\left(\frac{y^{3}}{27 x}\Big\vert {\ulamek{1}{3} \atop 0, \ulamek{1}{3}, \ulamek{2}{3}}\right) \\ \label{29/05-6}
& = \frac{\sqrt{3}}{2\pi x^{2/3}} G^{2, 0}_{0, 2}\left(\frac{y^{3}}{27 x}\Big\vert  {- \atop 0, \ulamek{2}{3}}\right) \\ \label{29/05-7}
& =  \frac{\sqrt{3}}{2\pi x^{2/3}} G\left([[\,\,\,], [\,\,\,]], [[0, \ulamek{2}{3}], [\,\,\,]], \ulamek{y^{3}}{27 x}\right) \\ \label{29/05-8}
& = \frac{1}{\sqrt{3} \pi x} y K_{\frac{2}{3}}\left(\frac{2 y^{\frac{2}{3}}}{3\sqrt{3 x}}\right),
\end{align}
where in Eq. \eqref{29/05-5} the simplification property of Meijer $G$ function was used and in Eq. \eqref{29/05-7} we applied the formula 8.4.23.1 of [\onlinecite{APPrudnikov-v3}], where $K_{\nu}(z)$ is the modified Bessel function of second kind of order $\nu$. \\ \\

\noindent
(b) \underline{Case $\alpha = 1/2$ and $\beta = 2/3$} \\
Here $\alpha\beta = 1/3$, $l=1$, $k=3$, $\alpha - 1/3 = 1/6$.
\begin{align}\label{1/06-1}
\mathcal{J}_{\frac{1}{3}, \frac{2}{3}}(x, y) & = \mathcal{L}^{-1}\left[p^{-\frac{1}{6}} \exp(-yp^{\frac{1}{3}}); x\right] = \frac{\sqrt{3}}{2\pi x^{5/6}} G^{3, 0}_{1, 3}\left(\frac{y^{3}}{27 x}\Big\vert {\ulamek{1}{6} \atop 0, \ulamek{1}{3}, \ulamek{2}{3}}\right) \\ \label{1/06-2}
& = \frac{\sqrt{3}}{2\pi x^{5/6}} G\left([[\,\,\,], [\ulamek{1}{6}]], [[0, \ulamek{1}{3}, \ulamek{2}{3}], [\,\,\,]], \ulamek{y^{3}}{27 x}\right).
\end{align}
The functions of Eqs. \eqref{1/06-1} and \eqref{1/06-2} can still be represented by better known special functions using the representation of Meijer $G$ function as a finite sum of generalized hypergeometric functions ${_{p}F_{q}}\left({(a_{p}) \atop (b_{q})}; x\right)$, see formula 8.2.2.3 of [\onlinecite{APPrudnikov-v3}], where $(a_{p})$ is a list of ``upper'' parameters and $(b_{q})$ is a list of ``lower'' parameters of ${_{p}F_{q}}$. An application of the formula 8.2.2.3 of [\onlinecite{APPrudnikov-v3}] to Eq. \eqref{1/06-2} gives directly
\begin{align}\label{1/06-3}
\mathcal{J}_{\frac{1}{2}, \frac{2}{3}}(x, y) & = \frac{\Gamma(\ulamek{5}{6})}{2\pi x^{5/6}} {_{1}F_{2}}\left({\ulamek{5}{6} \atop \ulamek{1}{3}, \ulamek{2}{3}}; \frac{y^{3}}{27 x}\right) + \frac{y^{\ulamek{1}{3}}}{6 \Gamma(\ulamek{5}{6}) x^{7/6}} {_{1}F_{2}}\left({\ulamek{7}{6} \atop \ulamek{2}{3}, \ulamek{4}{3}}; \frac{y^{3}}{27 x}\right) \nonumber \\
& - \frac{y^{2}}{4\sqrt{\pi} x^{3/2}} {_{1}F_{2}}\left({\ulamek{3}{2} \atop \ulamek{4}{3}, \ulamek{5}{3}}; \frac{y^{3}}{27 x}\right),
\end{align}
which also agrees with the formula 2.2.1.6 of [\onlinecite{APPrudnikov-v5}] for the parameter $\nu = 1/6$, see however [\onlinecite{CORR}], for its corrected form. \\

\noindent
(c) \underline{Case $\alpha = \beta = 1/2$} \\
Here $\alpha \beta = 1/4$, $l = 1$, $k=4$, $\alpha - 1/4 = 1/4$.
\begin{align}\label{1/06-4}
\mathcal{J}_{\frac{1}{2}, \frac{1}{2}}(x, y) & = \mathcal{L}^{-1}\left[p^{-1/4} \exp\left(-y p^{\frac{1}{4}}\right); x\right] = \frac{2}{(2\pi)^{3/2} x^{3/4}} G^{4, 0}_{1, 4}\left(\frac{y^{4}}{4^{4} x}\Big\vert {\ulamek{1}{4} \atop 0, \ulamek{1}{4}, \ulamek{1}{2}, \ulamek{3}{4}}\right) \\ \label{1/06-5}
& = \frac{2}{(2\pi)^{3/2} x^{3/4}} G\left([[\,\,\,], [\ulamek{1}{4}]], [[0, \ulamek{1}{4}, \ulamek{1}{2}, \ulamek{3}{4}], [\,\,\,]], \ulamek{y^{4}}{256 x}\right) \\ 
& = \frac{\Gamma(\ulamek{3}{4})}{\sqrt{2}\pi x^{3/4}} {_{0}F_{2}}\left({- \atop \frac{1}{4}, \frac{1}{2}}; -\frac{y^{4}}{256 x}\right) - \frac{y^{2}}{8 \Gamma(\ulamek{3}{4}) x^{5/4}} {_{0}F_{2}}\left({- \atop \ulamek{3}{4}, \ulamek{3}{2}}; -\frac{y^{4}}{256 x}\right) \nonumber \\ \label{1/06-7}
& + \frac{y^{3}}{12 \sqrt{\pi} x^{3/2}} {_{0}F_{2}}\left({- \atop \ulamek{5}{4}, \ulamek{7}{4}}; -\ulamek{y^{4}}{256 x}\right),
\end{align}
where in obtaining Eq. \eqref{1/06-7} the formula 8.2.2.3 of [\onlinecite{APPrudnikov-v3}] was again used.

\section{Discussion and conclusions}

The main objective of this work is the formulation and the proof of two new properties of the Laplace transform, s. Theorem 1, Eqs. \eqref{21Feb14-15} and \eqref{21Feb14-16} involving integration with kernels expressible through one-sided L\'{e}vy stable distributions $g_{\alpha}(x)$, $0 < \alpha < 1$. These results were obtained by first noting that two already known relations, Eqs. \eqref{21Feb14-3} and \eqref{21Feb14-4}, could be reworded in terms of L\'{e}vy-Smirnov function $g_{1/2}(x)$ of Eq. \eqref{21Feb14-6}. This observation allowed further extension for arbitrary $\alpha$ by choosing appropriate kernels $M_{\alpha}(t, x)$ and $N_{\alpha}(t, x)$ of Eqs. \eqref{21Feb14-9} and \eqref{21Feb14-10}, respectively, thereby permitting to prove two Laplace transform formulas, Eqs. \eqref{21Feb14-13c} and \eqref{21Feb14-14f}. Since $g_{\alpha}(x)$ are explicitly known for $\alpha$ rational [\onlinecite{KAPenson10}] that gives a wealth of novel Laplace inversion formulas. Subsequently the integration kernels $M_{\alpha}(t, x)$ and $N_{\alpha}(t, x)$ were shown to possess the transitivity  (or a kind of a reproducing property) with respect to the index $\alpha$, under integration of convolution type. This can be clearly seen from Eqs. \eqref{24/05-3} and \eqref{24/05-7} which, in the limit $\alpha\to 1$, become proportional to Mellin convolution of two L\'{e}vy stable distributions (with the caveat that the value $\alpha = 1$ is outside the range of our interest here). 

It turns out that even more general kernel correlation function, defined by Eqs. \eqref{24/05-10} and \eqref{24/05-18} can be exactly evaluated but for rational $\alpha$ and $\beta$ only, again using the Laplace transform technique. Observe in passing that Eq. \eqref{24/05-17} as a function of $p$, can be cast in the form of Weibull distribution [\onlinecite{NBalakrishnan03}], signaling a link between L\'{e}vy stable and extreme values distributions, which is a subject of recent publications, see [\onlinecite{TSimon14}] and [\onlinecite{KAPenson14}]. Note that the calculation of closed form of the Laplace transform in Eq. \eqref{24/05-10}, but with $y$ as integration variable, does not appear to be possible.

The kernels appearing in Eqs. \eqref{21Feb14-9} and \eqref{21Feb14-10} are intimately related to the fractional Fokker-Planck equation. In fact the kernel $N_{\alpha}(t, x)$ appears in the work of E. Barkai, see Eq. (16) in [\onlinecite{Bark}], whereas the kernel $M_{\alpha}(t, x)$ was used in Eq. (25) of Ref. [\onlinecite{FFP}], in connection with partial differential equations in which fractional derivatives are acting on the spatial coordinates.

The field of applications of L\'{e}vy stable laws is huge. It ranges from theories of random matrices [\onlinecite{ZBurda02}], description of optical properties of nanocrystals [\onlinecite{GDattoli14}] to predator search behavior in marine biology [\onlinecite{DWSims08}]. Large number of the other examples can be quoted. Various calculational aspects of L\'{e}vy stable laws are critically reviewed in [\onlinecite{TKPogany14}]. Theoretical and experimental aspects are reviewed in [\onlinecite{APityatinska05}] and [\onlinecite{WAWoyczynski01}], respectively.

\section*{Acknowledgments}
We thank the anonymous referee for informing us about the Ref. [\onlinecite{Stankovic}], unknown to us at the time of writing this paper. In [\onlinecite{Stankovic}] our Eq. \eqref{21Feb14-16} was obtained using the method of complex analysis.

The authors acknowledge support from the PHC Polonium, Campus France, project no. 28837QA and the bilateral collaboration project between PAN (Poland) and CNRS (France). KG thanks support from MNiSW, "Iuventus Plus 2015-2016", program no. IP2014 013073.

%------------------------------------------


\begin{thebibliography}{39}
\bibitem{CAS} We have extensively used Maple$^{\text{\textregistered}}$ and Mathematica$^{\text{\textregistered}}$ in this work.
\bibitem{CORR} We would like to correct a misprint in the formula 2.2.1.6 p. 51 of [\onlinecite{APPrudnikov-v5}]: the third hypergeometric function in this formula should be ${_{1}F_{2}}\left({\ulamek{5}{3} - \nu \atop \ulamek{4}{3}, \ulamek{5}{3}}; \frac{a^{3}}{27 x}\right)$.
\end{thebibliography}
\end{document}